\documentclass[12pt]{article}

\usepackage{amssymb}
\usepackage{citesort}
\usepackage{amsmath}
\usepackage{url}
\usepackage{color}

\newcommand{\db}{de$\,$Broglie}

\def\de{\delta}

\def\na{\nabla}
\def\la{\langle}
\def\ra{\rangle}
\def\pa{\partial}
\def\fr{\frac}

\def\al{\alpha}

\begin{document}
\vspace*{1.0cm}
\noindent
{\bf
{\large
\begin{center}
A minimalist pilot-wave model for quantum electrodynamics
\end{center}
}
}

\vspace*{.5cm}
\begin{center}
W.\ Struyve, H.\ Westman
\end{center}

\begin{center}
Perimeter Institute for Theoretical Physics \\
31 Caroline Street North, Waterloo, Ontario N2L 2Y5, Canada \\
E--mail: wstruyve@perimeterinstitute.ca, hwestman@perimeterinstitute.ca
\end{center}

\begin{abstract}
\noindent
We present a way to construct a pilot-wave model for quantum electrodynamics. The idea is to introduce beables corresponding only to the bosonic degrees of freedom and not to the fermionic degrees of freedom of the quantum state. We show that this is sufficient to reproduce the quantum predictions. The beables will be field beables corresponding to the electromagnetic field and they will be introduced in a similar way to that of Bohm's model for the free electromagnetic field. Our approach is analogous to the situation in non-relativistic quantum theory, where Bell treated spin not as a beable but only as a property of the wavefunction. After presenting this model we also discuss a simple way for introducing additional beables that represent the fermionic degrees of freedom.
\end{abstract}

\bibliographystyle{unsrt}

\section{Introduction}
In his seminal paper in 1952, Bohm presented a pilot-wave interpretation for the free electromagnetic field \cite{bohm522}. In this pilot-wave interpretation, the additional variable, also called beable, which together with the quantum state makes up the description of a system, is a field. Similar pilot-wave models can be constructed for other bosonic fields, like for example the scalar field. For reviews of these models see \cite{kaloyerou85,kaloyerou94,valentini92,valentini04,bohm872,bohm93,holland,holland93,struyve05,struyve071}. 

For fermionic quantum fields the introduction of field beables does not appear so straightforward. There is an attempt by Valentini \cite{valentini92,valentini96}, who took anti-commuting fields as beables, and an attempt by Holland \cite{holland,holland881}, who took a field of angular variables as beables. However, Valentini's model is problematic since it fails to reproduce the quantum probabilities \cite{struyve05,struyve071}. In addition, Holland's model might not be viable, since it is unclear whether measurement results get recorded in the beables in this model (for a detailed analysis, see \cite{struyve071}). Future work might put these models on a more solid footing.

So far, particle beables seem more successful for fermionic quantum field theory. Bell presented a model for quantum field theory on a lattice \cite{bell872}, where the beables are the fermion numbers at each lattice point. Bell's model differs from the usual pilot-wave program in the fact that it is indeterministic. However, Bell expected that the indeterminism would disappear in the continuum limit. A continuum model put forward by Colin \cite{colin031,colin032,colin033}, which was further developed by Colin and Struyve \cite{colin07}, seems to confirm Bell's expectation. On the other hand, D\"urr {\em et al.} have developed a continuum version of Bell's model which is stochastic \cite{durr02,durr031,durr032,tumulka03,durr04}. The fact that there are two generalizations of Bell's lattice model for the continuum originates in a different reading of Bell's work \cite{tumulka06,colin07}.

In this paper we present an alternative approach to a pilot-wave model for quantum field theory. We will argue that it is sufficient to introduce beables corresponding only to the bosonic degrees of freedom. No beables need to be introduced corresponding to the fermionic degrees of freedom. We illustrate this idea for quantum electrodynamics. The beables will be field beables corresponding to the electromagnetic field and they will be introduced in a similar way to that of Bohm's model for the free electromagnetic field. In this way we obtain a deterministic pilot-wave model. The strategy of not associating beables to all the degrees of freedom of the quantum state has been exploited before in some pilot-wave models. There is for example Bell's model for non-relativistic spin-1/2 particles, where no beables are associated with the spin degrees of freedom. 

In order to appreciate our minimalist proposal at this stage, it might be useful to keep in mind that already classically, charge distributions leave an imprint on, and can be inferred from, the classical radiation field. In the same way, we will argue that the beables corresponding to the electromagnetic field (and more precisely the radiation field) records and displays outcomes of measurements, and, more generally, contains an image of the everyday classical world.

After presenting this minimalist model, we also present a simple way of amending the model with beables for fermionic degrees of freedom. However, we want to emphasize that, for the purpose of reproducing the quantum predictions, such extra beables are not necessary.

Our pilot-wave model is not Lorentz covariant at the fundamental level. However, since the models reproduce the predictions of standard quantum field theory, Lorentz covariance is regained at the statistical level. This is discussed in detail in the context of other pilot-wave models, see e.g.\ \cite{valentini92,bohm93,holland}. Similar discussions apply to our pilot-wave models for quantum electrodynamics. In \cite{tumulka06} one can find a good review of attempts to construct Lorentz covariant models. See also \cite{valentini97} for an argument that the fundamental symmetry of pilot-wave theories is Aristotelean rather than Lorentzian.

In the next section we start with recalling the pilot-wave theory for non-relativistic quantum systems that was originally presented by \db\ and Bohm \cite{debroglie28,bohm521,bohm522}. We will thereby reconsider in detail how the pilot-wave theory reproduces the predictions of standard quantum theory. When we present our pilot-wave model for quantum field theory, we will use similar arguments in order to show that our model reproduces the standard quantum predictions. Then, in Section \ref{spin1/2}, we review Bell's model for non-relativistic spin-1/2 particles. This model serves as a simple illustration that it is not necessary to associate beables to every degree of freedom for the wavefunction in order to reproduce the quantum predictions. In Section \ref{themodel}, we present our model for quantum electrodynamics and show how it reproduces the standard quantum predictions. Finally, in Section \ref{beablesforfermions}, we present a simple way of introducing additional beables for the fermionic degrees of freedom.

\section{Pilot-wave theory of \db\ and Bohm}
In the pilot-wave theory for non-relativistic quantum systems of \db\ and Bohm \cite{debroglie28,bohm521,bohm522}, the complete description of a quantum system is provided by its wavefunction $\psi$, which satisfies the non-relativistic Schr\"odinger equation, and by particles which have definite positions ${\bf x}_k$, with $k=1,\dots,N$, at all times. The possible trajectories for the particles are solutions to the guidance equations 
\begin{equation}
\fr{d {\bf x}_k}{dt}= \frac{1}{2i\hbar m_k|\psi|^2} \left( \psi^* {\boldsymbol {\nabla}}_k \psi -\psi{\boldsymbol {\nabla}}_k \psi^*   \right)   = \frac{1}{m_k} {\boldsymbol {\nabla}}_k S\,,
\label{1}
\end{equation}
where $\psi=|\psi|\exp(iS/\hbar)$. 

If we consider an ensemble of systems that are all described by the same wavefunction, then the positions of the particles will have a certain distribution. In the special case in which this distribution equals $|\psi|^2$, the distribution is called the {\it equilibrium distribution} \cite{valentini911,durr92}. We have the special property that, if the distribution equals the equilibrium distribution, i.e.\ $|\psi({\bf x}_1,\dots,{\bf x}_N,t_0)|^2$, at one time $t_0$, then the distribution will equal the equilibrium distribution, i.e.\  $|\psi({\bf x}_1,\dots,{\bf x}_N,t)|^2$, at all times $t$. This property of the distribution is also called {\it equivariance} \cite{durr92} and follows from the fact that the density $|\psi|^2$ satisfies the continuity equation
\begin{equation}
\frac{\partial |\psi|^2}{\partial t} + \sum^N_{k=1}{\boldsymbol {\nabla}}_k \cdot \left(  \frac{{\boldsymbol {\nabla}}_k S}{m_k} |\psi|^2   \right) =0 \,,
\label{2}
\end{equation}
which is itself a consequence of the Schr\"odinger equation.{\footnote{Actually an easy way to rederive the pilot-wave theory consists in considering the continuity equation for the probability density $|\psi|^2$ and by postulating a velocity field for the beables that is given by the ratio of the probability current and the probability density. This is the technique we will apply when constructing our pilot-wave model for quantum electrodynamics. Note that neither \db\ or Bohm derived the pilot-wave theory in this way. On the other hand, this derivation was often applied by Bell.}}

Having the equilibrium distribution for ensembles is a key ingredient in showing that the pilot-wave theory reproduces the predictions of standard quantum theory. As we will explain below, it will guarantee that we will recover the Born rule. Possible justifications for the equilibrium distribution are studied in \cite{valentini911,valentini92,valentini04,valentini042,durr92,bohm93}.

Another key ingredient of the pilot-wave theory is that wavefunctions representing macroscopic systems that occupy different regions in physical 3-space have negligible overlap. This will guarantee that measurement results get recorded and displayed in the position configurations. It will also give rise to an effective collapse in measurement situations.

The analysis of how the pilot-wave interpretation for non-relativistic quantum systems reproduces the quantum predictions was first given in \cite{bohm522} and has been repeated many times afterwards, see e.g.\ \cite{bell872,bohm871,struyve06} (for an extensive analysis see \cite{durr033}). Since standard quantum theory is primarily about statistical predictions about the outcomes of measurements, we can just consider a measurement situation and compare the standard quantum predictions with those of pilot-wave theory.

In standard quantum theory a measurement situation can be described as follows. The wavefunction $\psi$, describing the system under observation, together with the measurement device, macroscopic pointer and environment, evolves into a superposition $\sum_i c_i \psi_i$, where the different normalized $\psi_i$ correspond to the different possible outcomes of the measurement, i.e.\ the different $\psi_i$ represent the different possible macroscopic pointer configurations which indicate the different possible measurement outcomes, like for example a needle pointing in different directions. The wavefunction then collapses to one of the $\psi_i$, say $\psi_k$, and this with probability $|c_k |^2$. The state $\psi_k$ represents one of the possible macroscopic pointer configurations and hence also the corresponding outcome of the measurement.

In the pilot-wave theory the measurement situation is described as follows. As in standard quantum theory the wavefunction evolves into a superposition $\sum_i c_i \psi_i$, but this time there is no subsequent collapse. The measurement result does not get recorded in the wavefunction, but in the beable configuration. This can be seen as follows. Since the $\psi_i$ correspond to different macroscopic situations, like for example a needle pointing in different directions, they will have negligible overlap. This means that the actual beable configuration $({\bf x}_1,\dots,{\bf x}_n)$ will be in the support of only one of the wavefunctions $\psi_i$, say $\psi_k$. The configurations in the different supports  represent the different possible macroscopic situations. In this way the beable configuration displays the measurement outcome. Since in equilibrium, the probability for the beable configuration $({\bf x}_1,\dots,{\bf x}_n)$ to be in the support of $\psi_k$ is given by $|c_k |^2$, we also recover the quantum probabilities. 

Further, if the different wavefunctions $\psi_i$ stay approximately non-over\-lap\-ping at later times, which is usually guaranteed by decoherence effects, then as one can easily verify from the guidance equations, only one of the wavefunctions, namely $\psi_k$, will play a significant role in determining the velocity field of the actual beable configuration $({\bf x}_1,\dots,{\bf x}_n)$, from time $t$ onwards. Since the other wavefunctions $\psi_i$, $i \neq k$ play only a negligible role, they can be ignored in the future description of the behaviour of the beable configuration. This is called an effective collapse. The effective collapse explains the success of the ordinary collapse in standard quantum theory. Unlike the collapse in standard quantum theory the effective collapse is not one of the axioms of the theory, but a consequence of the theory. 

Note that the property that macroscopically distinct states are non-overlapping in configuration space is a special property of the position representation. The same states might not be non-overlapping in another representation. This is very important for the choice of beable since it is unclear how a pilot-wave model, in which macroscopically distinct states are overlapping in the corresponding configuration space reproduces the quantum predictions (see also \cite{bacciagaluppi04}). For example, in Holland's model for fermionic fields \cite{holland,holland881} it is unclear whether macroscopically distinct states are non-overlapping in the configuration space of these angular variables and this might undermine the pilot-wave model \cite{struyve071}.

\section{Bell's model for non-relativistic spin-1/2 particles}\label{spin1/2}
In \cite{bell66,bell71}, Bell proposed a pilot-wave model for non-relativistic spin-1/2 particles by introducing beables only for the position degrees of freedom of the wavefunction and not for the spin degrees of freedom. Let us see how this works. Consider for simplicity just a single particle.

The wavefunction $\psi_a({\bf x},t)$ for a non-relativistic spin-1/2 particle satisfies the Pauli equation
\begin{equation}
i\hbar\frac{\pa \psi_a}{\pa t}=-\frac{\hbar^2}{2m}\left(\boldsymbol{\nabla}-\frac{ie}{\hbar c}{\bf A}\right)^2\psi_a + \sum_{b} \mu{{\widehat{\boldsymbol \sigma}}}_{ab}\cdot{\bf B} \psi_b+V\psi_a\,,
\label{3.6}
\end{equation}
with ${\bf A}$ the electromagnetic vector potential, ${\bf B}=\boldsymbol{\nabla}\times{\bf A}$ the corresponding magnetic field and $V$ an additional scalar potential. $\mu$ is the magnetic moment. 

In Bell's model the only beable is a position ${\bf x}$. The guidance equation is obtained by considering the continuity equation for the position density 
\begin{equation}
\rho^{\psi}({\bf x},t) = \sum_{a}|\psi_a ({\bf x},t)|^2 \,,
\label{3.8}
\end{equation}
which reads
\begin{equation}
\frac{\pa \rho^{\psi}}{\pa t} +\boldsymbol{\nabla} \cdot {\bf j}^{\psi}=0\,,
\label{4}
\end{equation}
with
\begin{equation}
{\bf j}^{\psi}=\sum_{a}\left(\fr{\hbar}{2mi}\left(\psi^{*}_a\boldsymbol{\nabla}\psi_a - \psi_a \boldsymbol{\nabla}\psi^{*}_a\right)-\fr{e}{mc}{\bf A}\psi^{*}_a\psi_a\right)\,.
\label{5}
\end{equation}
 The guidance equation is then given by
\begin{equation}
\fr{d {\bf x}}{dt}=\fr{{\bf j}^{\psi}}{\rho^{\psi}}\,.
\label{6}
\end{equation}
This guidance equation ensures that the equilibrium density $\rho^{\psi}({\bf x},t)$ is preserved under the dynamics, so it is equivariant.

This model is easily generalized to many particles. No beables are introduced for the spin degrees of freedom of the wavefunction. However, since results of measurements are generally recorded in ``positions of things'', like positions of macroscopic needle pointers, the same analysis as in the pilot-wave model of \db\ and Bohm for non-relativistic systems can be applied in order to show that the model reproduces the predictions of standard quantum theory. 

In Bell's model the only beables are particle positions. While there is no need to introduce additional beables for spin, nor for any other observables, there exist models which have such beables, like for example the models of Bohm {\em et al.}\ \cite{bohm551,bohm552} and Holland \cite{holland881,holland882,holland}. (See also \cite{tumulka062} for further comments.)

\section{Pilot-wave model for QED}\label{themodel}
In this section we present our pilot-wave model for quantum electrodynamics (QED). We start with recalling some features of the standard formulation of QED. The pilot-wave model in presented in Section \ref{pilot-wavemodel}. In Section \ref{quantumpredictions}, we will then explain how our pilot-wave model reproduces the quantum predictions.{\footnote{In this section we use units in which $\hbar = c =1$.}}

\subsection{Standard formulation of QED}\label{standard formulation}
\subsubsection{Field operators}
We start with the formulation of QED in the Coulomb gauge, which can be found in e.g.\ \cite[pp.\ 346-350]{weinberg95}.  This formulation is in terms of the operators ${\widehat \psi},{\widehat \psi}^{\dagger},{\widehat {\bf A}}^T$ and ${\widehat {\boldsymbol \Pi}}^T$. The operators ${\widehat \psi}$ and ${\widehat \psi}^{\dagger}$ are the Dirac field operators and the operators ${\widehat {\bf A}}^T$ and ${\widehat {\boldsymbol \Pi}}^T$ are respectively the transverse electromagnetic field operator and the transverse momentum field operator, i.e.\ we have ${\boldsymbol \na} \cdot {\widehat {\bf A}}^T ={\boldsymbol \na} \cdot {\widehat {\boldsymbol \Pi}}^T = 0 $. 

The anti-commutation relations of the fermionic field operators are given by 
\begin{equation}
{\{{\widehat \psi}_a({\bf x}), {\widehat \psi}^{\dagger}_b({\bf y})\} = \delta_{ab} \delta({\bf x} - {\bf y})}\,,
\label{12}
\end{equation}
where the other anti-commutation relations are zero. The commutation relations of the bosonic field operators read
\begin{equation}
[{\widehat A}^T_i({\bf x}),{\widehat \Pi}^T_{j}({\bf y})] = i \left(\delta_{ij} -  \frac{\partial_i \partial_j }{\nabla^2}\right)\delta({\bf x} - {\bf y}) \,,
\label{6.002}
\end{equation}
where the other commutation relations are zero. The fermionic and bosonic operators mutually commute.

The Hamiltonian reads
\begin{equation}
{\widehat H} = {\widehat H}_B + {\widehat H}_F + {\widehat H}_I + {\widehat V}_C\,,
\label{7}
\end{equation}
where
\begin{equation}
{\widehat H}_B = \fr{1}{2} \int d^3 x \left( {\widehat {\boldsymbol \Pi}}^T \cdot {\widehat {\boldsymbol \Pi}}^T - {\widehat {\bf A}}^T \cdot \nabla^2  {\widehat {\bf A}}^T    \right)
\label{6.001}
\end{equation}
is the free Hamiltonian for the electromagnetic field, ${\widehat H}_F$ is the free Hamiltonian for the Dirac field
\begin{equation}
{\widehat H}_F = \int d^3 x {\widehat \psi}^{\dagger}  \left( -i {\boldsymbol \al} \cdot  {\boldsymbol \na}  \right){\widehat \psi}\,,     
\label{8}
\end{equation}
${\widehat H}_I$ is the interaction Hamiltonian
\begin{equation}
{\widehat H}_I = - \int d^3 x  {\widehat {\bf A}}^T \cdot {\widehat {\bf j}}\,, 
\label{9}
\end{equation}
and ${\widehat V}_C$ is the Coulomb energy
\begin{equation}
{\widehat V}_C = \frac{1}{2}\int d^3 x d^3 y \frac{{\widehat j}^0({\bf x}){\widehat j}^0({\bf y})}{4\pi |{\bf x} -{\bf y}|}\,.
\label{10}
\end{equation}
The operator ${\widehat j}^\mu$ is the Dirac charge current density operator
\begin{equation}
{\widehat j}^\mu = e {\widehat \psi}^{\dagger} \gamma^0 \gamma^\mu  {\widehat \psi} =  e \left( {\widehat \psi}^{\dagger} {\widehat \psi} ,   {\widehat \psi}^{\dagger}  {\boldsymbol \al} {\widehat \psi}     \right)\,.
\label{11}
\end{equation}

For future purposes we also introduce the magnetic field operator ${\widehat {\bf B}} = {\boldsymbol \nabla } \times {\widehat {\bf A}}^T$ and the transverse part of the electric field operator ${\widehat {\bf E}}^T = - {\widehat {\boldsymbol \Pi}}^T $. The longitudinal part of the electric field operator is related to the charge density operator by the Gauss law $ {\boldsymbol \nabla } \cdot {\widehat {\bf E}}^L ={\widehat j}^0$.

\subsubsection{Representation for the field operators}
We take the Hilbert space to be the product of a bosonic Hilbert space and a fermionic Hilbert space. The electromagnetic field operators and the Dirac field operators respectively act on the bosonic and fermionic Hilbert space. In order to find a representation for the product Hilbert space we first consider separately a representation for the bosonic and for the fermionic Hilbert space.

A representation for the electromagnetic field operators can be found by considering the Fourier expansion
\begin{eqnarray}
{\widehat {\bf A}}^T({\bf x}) &=& \frac{1}{(2\pi)^{3/2}} \sum^2_{l=1} \int d^3 k e^{i {\bf k} \cdot {\bf x}} {\boldsymbol \varepsilon}^l({\bf k}) {\widehat q}_l({\bf k})\,, \label{6.0021}\\  
{\widehat {\boldsymbol \Pi}}^T({\bf x})  &=&  \frac{1}{(2\pi)^{3/2}} \sum^2_{l=1} \int d^3 k e^{-i {\bf k} \cdot {\bf x}} {\boldsymbol \varepsilon}^l({\bf k}) {\widehat \pi}_l({\bf k})\,.
\label{6.003}
\end{eqnarray}
Here $ {\widehat q}_l$ and ${\widehat \pi}_l$ are operators in momentum space which satisfy the commutation relations
\begin{equation}
[{\widehat q}_l({\bf k}) ,{\widehat \pi}_{l'}({\bf k}') ] = i \delta_{ll'}\delta({\bf k} - {\bf k}') \,, \quad [{\widehat q}_l({\bf k}) ,{\widehat q}_{l'}({\bf k}') ] = [{\widehat \pi}_l({\bf k}),{\widehat \pi}_{l'}({\bf k}')]= 0\,.
\label{6.00301}
\end{equation}
The vectors ${\boldsymbol \varepsilon}^l({\bf k})$, $l=1,2$ are two real, orthogonal polarization vectors, which we choose to obey the following relations
\begin{equation}
{\bf k} \cdot {\boldsymbol \varepsilon}^l({\bf k}) = 0\,, \quad \sum^2_{l=1} {\varepsilon}^l_i({\bf k}) {\varepsilon}^l_j({\bf k})= \delta_{ij} - \frac{k_i k_j}{k^2}\,, \quad {\boldsymbol \varepsilon}^l({\bf k}) = {\boldsymbol \varepsilon}^l(-{\bf k})\,.
\label{6.004}
\end{equation}
From the last relation and the fact that ${\widehat {\bf A}}^T$ and ${\widehat {\boldsymbol \Pi}}^T$ are Hermitian, we have that ${\widehat q}_l({\bf k}) = {\widehat q}^{\dagger}_l(-{\bf k})$ and ${\widehat \pi}_l({\bf k})={\widehat \pi}^{\dagger}_l(-{\bf k})$. One can easily show that the commutation relations for the field operators ${\widehat q}_l({\bf k})$ and ${\widehat \pi}_l({\bf k})$ give the commutation relations (\ref{6.002}) for ${\widehat {\bf A}}^T({\bf x})$ and ${\widehat {\boldsymbol \Pi}}^T({\bf x})$.

At this point we can introduce the eigenstates $ | q_1,q_2 \rangle$ of the operators ${\widehat q}_l$, $l=1,2$, where ${\widehat q}_l({\bf k}) = {\widehat q}^{\dagger}_l(-{\bf k})$ and $ {\widehat q}_{l}({\bf k})  | q_1,q_2 \rangle  =  q_{l}({\bf k})  | q_1,q_2 \rangle$. We further assume the functions $q_{l}({\bf k})$ to be smooth. With the states $| q_1,q_2 \rangle$ as a basis of the bosonic Hilbert space, we obtain the functional Schr\"odinger representation for the bosonic field operators, cf.\ equation (\ref{16.1}) below.{\footnote{For a detailed treatment of quantum field theory in the functional Schr\"odinger picture, see Hatfield \cite{hatfield91}.}}

For the fermionic field operators we will not choose an explicit representation because we will integrate out the fermionic degrees of freedom. We just assume that the fermionic Hilbert space is spanned by some orthogonal states $|f\rangle$, labelled by an index $f$ which can be discrete or continuous. 

In the product basis $| q_1,q_2 \rangle \otimes | f \rangle=| q_1,q_2, f \rangle$ of the product Hilbert space, the field operators have matrix components
\begin{eqnarray}
\langle q_1,q_2 , f | {\widehat \psi}({\bf x}) | q'_1,q'_2, f' \rangle  &=&  \delta (q_1 - q'_1)\delta (q_2 - q'_2) \langle  f | {\widehat \psi} ({\bf x})|  f' \rangle  \,, \nonumber\\
\langle q_1,q_2 , f | {\widehat \psi}^{\dagger}({\bf x}) | q'_1,q'_2, f' \rangle  &=&  \delta (q_1 - q'_1)\delta (q_2 - q'_2) \langle  f | {\widehat \psi}^{\dagger} ({\bf x})|  f' \rangle \,, \nonumber\\
\langle q_1,q_2 , f | {\widehat q}_l({\bf k}) | q'_1,q'_2, f' \rangle  &=& q_l({\bf k})  \delta (q_1 - q'_1)\delta (q_2 - q'_2)  \delta_{ff'}  \,, \nonumber\\  
\langle q_1,q_2 , f | {\widehat \pi}_l({\bf k})| q'_1,q'_2, f' \rangle  &=& - i \fr{\de }{\de q_l({\bf k})}  \delta (q_1 - q'_1)\delta (q_2 - q'_2)  \delta_{ff'}   \,. 
\label{16.1}
\end{eqnarray}
The components of the Hamiltonian will be written as
\begin{equation}
\langle q_1,q_2 , f | {\widehat H} | q'_1,q'_2, f' \rangle = {\widehat H}_{ff'} (q,-i\delta / \delta q) \delta (q_1 - q'_1)\delta (q_2 - q'_2) \,.
\label{16.2}
\end{equation}
For example, the bosonic part of the Hamiltonian hereby reads
\begin{equation}
\left( {\widehat H}_B \right)_{ff'}(q, -i\delta / \delta q) = \delta_{ff'} \frac{1}{2} \int d^3 k \left( - \frac{\de^2}{\de q^*_{l}({\bf k}) \de q_{l}({\bf k}) } + k^2  q^*_{l}({\bf k}) q_{l}({\bf k})  \right)\,.
\label{16.3}
\end{equation}

A state $| \Psi (t)\rangle$ has the expansion coefficients $\Psi_f(q_1,q_2,t)=\la  q_1,q_2 , f |\Psi(t) \ra$. So the expansion coefficients are wavefunctionals on the configuration space of fields, just as in the case of the free electromagnetic field \cite{bohm522}, with the difference that in this case the wavefunctionals carry an extra label $f$ which represents the fermionic degrees of freedom. Note the analogy with spin, where the wavefunction lives on ordinary configuration space and carries a spin-index. 

Using the notation introduced in equation (\ref{16.2}) we find that $\Psi_f(q_1,q_2,t)$ satisfies the functional Schr\"odinger equation
\begin{equation}
i \frac{ \pa \Psi_f(q_1,q_2,t)}{\pa t} = \sum_{f'} {\widehat H}_{ff'} (q,-i\delta / \delta q)\Psi_{f'}(q_1,q_2,t)\,.
\label{16.4}
\end{equation}

Actually this Schr\"odinger equation is mathematically not well defined. As is common in quantum field theory it needs regularization. In work by Symanzik and L\"uscher it is shown that, in the case of a self-interacting scalar field, the functional Schr\"odinger picture can be made mathematically well-defined by introducing an extra renormalization constant \cite{symanzik81,luscher85}. This result can probably be generalized to the functional Schr\"odinger equation considered here. However, a simpler regularization consists in replacing continuous space by a bounded lattice or by assuming a bounded lattice in momentum space (the latter just means that a bounded space is assumed, so that the momentum integrals in the Fourier expansion are replaced by sums over momenta, and that an ultra-violet cut-off is assumed).

\subsection{Pilot-wave model}\label{pilot-wavemodel}
We can now construct a pilot-wave model with beables only for the electromagnetic field. More precisely, we introduce beables $q_1({\bf k})$ and $q_2({\bf k})$ for which the equilibrium density is given by 
\begin{equation}
\rho^{\Psi}(q_1,q_2,t) =  \sum_f |\Psi_{f}(q_1,q_2,t )|^2 \,.
\label{18}
\end{equation}
We formally regard $\rho^{\Psi}$ as a density with respect to the measure ${\mathcal D}q_1{\mathcal D}q_2$, where ${\mathcal D}q_i=\prod_{\bf k}d q_i({\bf k})$. However, since there exists no generalization of the Lebesgue measure to an infinite dimensional space, ${\mathcal D}q_1{\mathcal D}q_2$ is strictly speaking not defined as a measure. Although this issue needs further attention in the future, we will not address it here. We just want to mention that the problem disappears when we assume a regularization which makes the number of degrees of freedom finite.

Just as was done in Bell's model for non-relativistic spin-1/2 particles we consider the continuity equation for $\rho^{\Psi}$ in order to find a guidance equation for the field beables $q_1({\bf k})$ and $q_2({\bf k})$. Using the Schr\"odinger equation (\ref{16.4}) we find 
\begin{eqnarray}
\pa_t \rho^{\Psi}(q_1,q_2,t) &=& \sum_f \left( \pa_t \Psi^*_f(q_1,q_2,t ) \Psi_{f}(q_1,q_2,t )  + \Psi^*_f(q_1,q_2,t ) \pa_t  \Psi_{f}(q_1,q_2,t ) \right)  \nonumber\\
&=& \sum_{f,f'}  i\Bigg( \left(  {\widehat H}_{ff'}(q, -i\delta / \delta q)   \Psi_{f'}(q_1,q_2,t ) \right)^* \Psi_{f}(q_1,q_2,t )\nonumber\\
&&\mbox{} - \Psi^*_f(q_1,q_2,t ) {\widehat H}_{ff'}(q, -i\delta / \delta q)\Psi_{f'}(q_1,q_2,t ) \Bigg) \,.
\label{18.1}
\end{eqnarray}
In this expression only the kinetic part of the free Hamiltonian for the electromagnetic field survives. This kinetic term has components
\begin{equation}
\left( {\widehat H}^{{\textrm{kin}}}_B \right)_{ff'}(q, -i\delta / \delta q) = -\delta_{ff'}\frac{1}{2} \int d^3 k \frac{\de^2}{\de q^*_{l}({\bf k}) \de q_{l}({\bf k}) }\,.
\label{18.2}
\end{equation}
In this way we find the continuity equation
\begin{eqnarray}
\pa_t \rho^{\Psi}(q_1,q_2,t) &=& \frac{1}{2i}  \sum_f \sum^2_{l =1} \int d^3k \Bigg(  \Psi_{f}(q_1,q_2,t ) \frac{\de^2}{\de q^*_{l}({\bf k}) \de q_{l}({\bf k})} \Psi^*_f(q_1,q_2,t )\nonumber\\
&& \mbox{}- \Psi^*_f(q_1,q_2,t ) \frac{\de^2}{\de q^*_{l}({\bf k}) \de q_{l}({\bf k})} \Psi_{f}(q_1,q_2,t )   \Bigg)\nonumber\\
&=& - \sum^2_{l =1} \int d^3k \frac{\de}{\de q_{l}({\bf k})}  J^{\Psi}_{{l}}({\bf k};q_1,q_2,t)\,,
\label{19}
\end{eqnarray}
where
\begin{equation}
J^{\Psi}_{{l}}({\bf k};q_1,q_2,t) =\sum_f |\Psi_f(q_1,q_2,t )|^2 \frac{\delta S_f(q_1,q_2,t )}{\delta  q^*_l({\bf k})  } \,, \qquad  l=1,2\,,
\label{20}
\end{equation}
where we have used $\Psi_f= |\Psi_f|\exp(iS_f)$. 

We can now postulate the following guidance equation for the field beables
\begin{equation}
\fr{\pa q_{l}({\bf k},t)}{\pa t} = \frac{J^{\Psi}_{{l}}({\bf k};q_1,q_2,t)}{\rho^{\Psi}(q_1,q_2,t)} \,, \qquad  l=1,2\,.
\label{21}
\end{equation}

The fields $(q_1,q_2)$ live in momentum space. The ontology in physical 3-space is of course given by a transverse field ${\bf A}^T$ which is defined in terms of the fields $(q_1,q_2)$ through the relation \eqref{6.0021}. The field ${\bf A}^T$ further gives rise to a field ${\bf B}$ in physical space, which is given by 
\begin{equation}
{\bf B}({\bf x}) = {\boldsymbol \nabla} \times {\bf A}^T({\bf x}) =  \frac{i}{(2\pi)^{3/2}} \sum^2_{l=1} \int d^3 k e^{i{\bf k} \cdot {\bf x}} {\bf k}\times {\boldsymbol \varepsilon}^l ({\bf k}) q_l({\bf k})\,.
\label{6.011}
\end{equation}
The field ${\bf B}$ can be regarded as the beable corresponding to the magnetic field. (Just as the particle position is the position that is revealed in a quantum measurement of position in non-relativistic pilot-wave theory, the field beable ${\bf B}$ is the field that is revealed (locally) in a quantum measurement of the magnetic field.) 

For a solution $(q_1(t),q_2(t))$ to the guidance equation, we can also introduce a field ${\bf E}^T$, which is given by
\begin{equation}
{\bf E}^T({\bf x},t) = -\partial_t {\bf A}^T({\bf x},t) =  -\frac{1}{(2\pi)^{3/2}} \sum^2_{l=1} \int d^3 k e^{i {\bf k} \cdot {\bf x}} {\boldsymbol \varepsilon}^l({\bf k}) \partial_t q_l({\bf k},t)
\label{21.3}
\end{equation}
and which resembles the transverse part of the electric field when expressed in terms of the potential in the Coulomb gauge. Unlike to the field beable ${\bf B}$, we need the time evolution of the beables $(q_1,q_2)$ to construct the field ${\bf E}^T$. 
%In this way, the fields ${\bf B}$ and ${\bf E}^T$ take on roles analogous to position ${\bf x}(t)$ and momentum ${\bf p} = {\boldsymbol \nabla} S({\bf x}(t),t)$ in non-relativistic quantum theory. 
Alternatively one could have developed a pilot-wave model in which the basic beable corresponds to the transverse part of the electric field. 

Note that we introduced beables only for the transverse degrees of freedom of the vector potential. In particular, we did not introduce beables for the longitudinal degrees of freedom, nor for the scalar degrees of freedom of the electromagnetic field. This implies that we do not have beables corresponding with the charge density of the fermionic field. If one could introduce beables for the longitudinal degrees of freedom or scalar degree of freedom of the electromagnetic field, then they could be related to the charge density through the constraints $j^0 = - \nabla^2 A^0$ and $j^0 = {\boldsymbol \nabla} \cdot {\bf E}={\boldsymbol \nabla} \cdot {\bf E}^L$ \cite[pp.\ 346-350]{weinberg95}. However, in Section \ref{beablesforfermions} we will present an other way to introduce a beable corresponding to the charge density.

The pilot-wave model is presented here in the context of QED. But we see no problem in principle to construct a similar model, i.e.\ a model with beables corresponding solely to the electromagnetic field $A_\mu$, in the context of the standard model. However since in the standard model the electromagnetic field is unified with the weak interaction fields $W^+_\mu,W^-_\mu,Z_\mu$ it might be more natural to introduce beables also for these fields. In addition one could also introduce beables for the other bosonic fields, namely the strong interaction and Higgs fields.

\subsection{How the pilot-wave model reproduces the quantum predictions}\label{quantumpredictions}
In the case of the pilot-wave theory of \db\ and Bohm for non-relativistic quantum systems (which we will subsequently refer to as non-relativistic pilot-wave theory), we saw that having the equilibrium distribution for the beables, together with the fact that wavefunctions representing macroscopically distinct systems have negligible overlap, were the two key properties that allowed us to show that the quantum predictions are reproduced by the pilot-wave theory.

In our pilot-wave model for QED similar key properties are present. First of all we assume that the field beables are distributed according to the equilibrium distribution $\rho^{\Psi}$ at a certain time, so that by equivariance the beables are distributed according to the equilibrium distribution at all times. Secondly, we have that wavefunctionals representing macroscopically distinct states of the electromagnetic field are non-overlapping in the configuration space of fields (whereby it is understood that two wavefunctionals $\Psi^{(1)}_f(q_1,q_2,t)$ and $\Psi^{(2)}_f(q_1,q_2,t)$ have no overlap when their components have no mutual overlap, i.e.\ $\Psi^{(1)}_f(q_1,q_2,t)\Psi^{(2)}_{f'}(q_1,q_2,t) = 0 $ for all $(q_1,q_2)$, $f$ and $f'$). This is simply because two wavefunctionals that correspond to different classical electric and magnetic fields will yield approximately disjoint magnetic field distributions and hence approximately disjoint distributions of the fields ${\bf A}^T$ and $(q_1,q_2)$, so that the corresponding wavefunctionals have negligible overlap.

Note that in order for the wavefunctionals to be non-overlapping, they need not be centered around magnetic field configurations that are localized in different regions of physical space. It is for example sufficient that they are centered around magnetic fields which differ only in a certain region in physical space.{\footnote{In \cite{saunders99}, Saunders expressed a worry for using fields as beables. However, he made the unwarranted assumption that localized macroscopic bodies should be represented by localized field beables in order have a pilot-wave model that produces the familiar image of a classical world. As our model illustrates, there is no need for that.}}

Let us now consider a measurement situation in order to see how these key properties enable us to show that we reproduce the standard quantum predictions, even though there are no beables corresponding to the fermionic degrees of freedom. Suppose for definiteness that we perform a measurement on some quantum system, such that, on the level of the quantum state, the outcome of the measurement becomes correlated with the orientation of some macroscopic needle. In non-relativistic pilot-wave theory there are particle positions representing the needle, so that the orientation of the needle, and hence the outcome of the measurement, will be recorded and displayed in the particles positions. On the other hand, in our model for QED, there are no beables corresponding to the fermionic degrees of freedom. However, on the level of the quantum state, the direction of the macroscopic needle will get correlated with the radiation that is scattered off (or thermally emitted from, etc.)\ the needle. Because these states of radiation will be macroscopically distinct, they will be non-overlapping in the configuration space of fields and hence the outcome of the experiment will be recorded and displayed in the field beable corresponding to the radiation. 

The way the beable displays the outcome of the measurement is very natural: the beable will carry an image of a macroscopic needle, in a similar way as in classical mechanics, where a classical electromagnetic field that scatters off a needle will carry an image of that needle.{\footnote{One difference with classical mechanics is of course that in our pilot-wave model the basic beable corresponds only to the magnetic field and not to the electric field.}} For example, by performing a Fourier decomposition of the field in a particular spatial region and finding the spectrum of the radiation in a certain direction, one can form an image of the surroundings of the that region, and in particular of the orientation of the measurement needle.

It is clear that we have a similar situation in other measurement-like situations (results of measurements need not be conveyed to us by means of instrument needles, it could also be by means of a scintillation or computer screen): the results of measurement outcomes will become correlated with macroscopically distinct classical states of the electromagnetic field, so that we have a branching of the wavefunctional and a record of the measurement outcome in the field beable.  

In summary, non-relativistic pilot-wave theory reproduces the quantum predictions, since outcomes of measurements are recorded in the position beables associated to macroscopic objects. On the other hand, our model for QED reproduces the quantum predictions since outcomes of measurements are recorded in the electromagnetic field beable. In particular, the image of macroscopic objects (and, more generally, of the familiar macroscopic world) can be inferred from the electromagnetic field beable.

\section{Beables for fermions?}\label{beablesforfermions}
We have presented a pilot-wave model for QED. This model is minimalist since beables were introduced only for bosonic degrees of freedom of the quantum state and not for fermionic ones. However, one could also construct models which also include beables for the fermionic degrees of freedom. One possible way to do this is by introducing particle beables as in the work of Colin or D\"urr {\em et al.} 

Still, a more simple way to introduce beables for fermionic degrees of freedom is as follows. Suppose ${\widehat \varrho}({\bf x})$ is an operator, like e.g.\ the energy density of charge density operator. Then one can introduce a field beable $\varrho({\bf x},t)$, given by{\footnote{This expression corresponds to Holland's local expectation value for the operator ${\widehat \varrho}({\bf x})$, evaluated at the beable configuration \cite{holland}.}} 
\begin{equation}
\varrho({\bf x},t) = \frac{\sum_{ff'} \Psi^*_f(q_1,q_2,t) {\widehat \varrho}_{ff'}({\bf x})  \Psi_{f'}(q_1,q_2,t)}{\rho^{\Psi}(q_1,q_2,t)} \Bigg|_{(q_1,q_2)=(q_1(t),q_2(t))} \,,
\label{30}
\end{equation}
where ${\widehat \varrho}_{ff'}({\bf x}) = \la f|{\widehat \varrho}({\bf x})|f' \ra$ and where the expression on the right hand side is evaluated at the actual beable configuration $(q_1(t),q_2(t))$.

It is interesting to compare such a model with an extra beable $\varrho({\bf x})$ (or extra beables), to the minimalist model (which is in terms of the beables $(q_1(t),q_2(t))$ or ${\bf A}^T$). Suppose for example we add a ``charge density'' beable $\varrho_c({\bf x})$ to the model and let us, for definiteness, compare the models in the description of the double slit experiment for an electron. In the minimalist model there is no beable that represents the electron. It is only when the position measurement is performed (at the final screen), that the field beables $(q_1,q_2)$ or ${\bf A}^T$ will record and display the measurement result (in the way explained in the previous section). On the other hand, in the model with the additional beable, the field $\varrho_c({\bf x})$ represents the electron. This field will branch when the electron passes through the slits and will display interference effects afterwards. Only when an actual position measurement is performed, upon which the wavefunction of the electron will get entangled with the wavefunction of the measurement device and environment, will this field $\varrho_c({\bf x})$ become localized.

Thus there is no problem to add beables for the fermionic degrees of freedom. However, for the purpose of reproducing the quantum predictions, there is no need to do this, since the electromagnetic field beables already provide an image of the familiar classical world. This illustrates that there is nothing unique in the choice of beables, something which already was noted before in e.g.\ \cite{bell872,goldstein04,allori06}. In quantum equilibrium all these choices will be empirically indistinguishable, although this is not necessarily so in the case of quantum non-equilibrium.

\section{Conclusion}
In this paper we presented a deterministic pilot-wave model that reproduces the predictions of quantum electrodynamics. The model has the special feature that it contains no beables for fermionic degrees of freedom. At first sight it would seem absolutely necessary to include fermionic beables in order to give an account for the everyday classical world. However, we demonstrated that this is not the case: a beable corresponding to the radiation field is sufficient to give such an account. 

We also discussed how this minimalist model could be amended, in a rather simple way, with beables for fermionic degrees of freedom.  

These pilot-wave models, together with previous models (which were mentioned in the introduction), not only show that it possible to construct pilot-wave models for quantum field theory, they also illustrate that this can be done in a number of different ways.

\section{Acknowledgements}
We want to thank Sheldon Goldstein, Owen Maroney, Sebastiano Sonego, Rafael Sorkin and Antony Valentini for discussions. We are also grateful to the anonymous referees for valuable suggestions. WS is further grateful to Stijn De Weirdt for initial encouragement. Research at Perimeter Institute is supported in part by the Government of Canada through NSERC and by the Province of Ontario through MRI.

\end{document}